\documentclass[twocolumn]{aastex701}

\begin{document}

\title{Anisotropic Secondary Bias of Dark Matter Haloes in a $\Lambda$CDM Universe}

\author[orcid=0009-0000-5148-9457]{Qinglin Ma}
\affiliation{Department of Astronomy, Tsinghua University, Beijing 100084, China}
\email[show]{maql21@mails.tsinghua.edu.cn}  

\author[orcid=0000-0002-8711-8970]{Cheng Li} 
\affiliation{Department of Astronomy, Tsinghua University, Beijing 100084, China}
\email[show]{cli2015@tsinghua.edu.cn}
\correspondingauthor{Qinglin Ma and Cheng Li}

%% Use the \collaboration command to identify collaborations. This command
%% takes an optional argument that is either a number or the word "all"
%% which tells the compiler how many of the authors above the command to
%% show. For example "\collaboration[all]{(DELVE Collaboration)}" wil include
%% all the authors above this command.
%%
%% Mark off the abstract in the ``abstract'' 

\begin{abstract}

     Secondary bias is the dependence of halo clustering on properties beyond halo mass. Using the $z=0$ TNG300-1-Dark simulation, we study anisotropic secondary bias (ASB): the variation of secondary bias with direction relative to the halo major axis. We first use ordinary, orientation-averaged secondary bias (OSB) as a baseline to compare three environmental manifestations: halo-environment alignment, outer matter anisotropy, and tidal anisotropy. Matching tidal anisotropy suppresses much of the OSB, whereas matching halo-environment alignment or outer matter anisotropy does not. ASB behaves differently. It is weak for formation time, concentration, and triaxiality, but strong for both spin definitions and minor-to-major axis ratio; slowly rotating and more elongated haloes are more strongly aligned with filamentary structure. Matching halo-environment alignment substantially reduces the spin- and shape-dependent ASB signals, whereas matching tidal anisotropy or the outer matter axis ratio leaves them largely intact. Halo definition has little impact on ASB, yet strongly affects low-mass spin bias: including unbound particles can move dense-environment haloes with low bound-particle spin into the high all-particle-spin sample. These results clarify which clustering signals are associated with halo-environment alignment, matter anisotropy, or tidal anisotropy, and which are sensitive to halo definition.
     
     \end{abstract}
     
     %% Keywords should appear after the \end{abstract} command. 
     %% The AAS Journals now uses Unified Astronomy Thesaurus concepts:
     %% https://astrothesaurus.org
     %% You will be asked to selected these concepts during the submission process
     %% but this old "keyword" functionality is maintained in case authors want
     %% to include these concepts in their preprints.
     
     
     %% From the front matter, we move on to the body of the paper.
     %% Sections are demarcated by \section and \subsection, respectively.
     %% Observe the use of the LaTeX \label
     %% command after the \subsection to give a symbolic KEY to the
     %% subsection for cross-referencing in a \ref command.
     %% You can use LaTeX's \ref and \label commands to keep track of
     %% cross-references to sections, equations, tables, and figures.
     %% That way, if you change the order of any elements, LaTeX will
     %% automatically renumber them.
     %%
     %% We recommend that authors also use the natbib \citep
     %% and \citet commands to identify citations. The citations are
     %% tied to the reference list via symbolic KEYs. The KEY corresponds
     %% to the KEY in the \bibitem in the reference list below. 
     
     \section{Introduction} \label{sec:intro}
     
     %\latex\ \footnote{\url{http://www.latex-project.org/}} is a document markup
     %language that is particularly well suited for the publication of
     %mathematical and scientific articles \citep{lamport94}. 
     
     In the $\Lambda$CDM picture, dark matter haloes provide the link between large-scale structure and galaxy formation. Halo clustering is governed primarily by mass \citep{1996Mo}, but it also depends on secondary properties such as formation time, concentration, spin, and shape (\citealt{2004Sheth}; \citealt{2005Gao}; \citealt{2006Wechsler}; \citealt{2007Gao}; \citealt{2007Jing}; \citealt{2007Wang}; \citealt{2008Dalal}; \citealt{2008Li}; \citealt{2010Faltenbacher}; \citealt{2011Wang}; \citealt{2016Sunayama}; \citealt{2017Lazeyras}; \citealt{2018Salcedo}; \citealt{2019Sato}; \citealt{2019Han}; \citealt{2020Chen}; \citealt{2024Smith}). This effect is known as assembly bias \citep{2005Gao}, or more generally secondary bias \citep{2018Mao}.
     
     The physical origin of secondary bias has often been linked to the tidal field (\citealt{2007Wang}; \citealt{2009Hahn_tidalforce}; \citealt{2011Wang}; \citealt{2015Shi}). In particular, tidal anisotropy near a halo can explain several secondary-bias signals (\citealt{2018Paranjape_tidalaniso}; \citealt{2019Ramakrishnan_tidalaniso}) and connects naturally to cosmic-web effects on halo accretion, structure, and clustering (\citealt{2017Borzyszkowski}; \citealt{2018Musso_EPScosmicweb_AB}; \citealt{2021Akitsu}; \citealt{2020Morinaga}; \citealt{2024Montero_Cosmicwebbias}; \citealt{2024Balaguera_assemblybiasdiffz}). At the same time, halo shapes are not randomly oriented with respect to the cosmic web. Halo major axes preferentially align with filaments, especially at higher mass (\citealt{2005Bailin}; \citealt{2008Paz}; \citealt{2009Faltenbacher}; \citealt{2011Paz}; \citealt{2018Ganeshaiah}; see also \citealt{2015Joachimi}; \citealt{2015Kiessling}). Environmental anisotropy is therefore imprinted on several connected scales, from the large-scale tidal field to the matter distribution around individual haloes and ultimately to their shapes, spins, and orientations.
     
     These connections involve three related but distinct environmental manifestations. The first is halo-environment alignment: the degree to which the halo major axis is aligned with the surrounding density or tidal field. This describes how the halo orientation is coupled to the surrounding density or tidal structure. The second is the anisotropy amplitude of the outer matter distribution around the halo. The third is local tidal anisotropy, describing the anisotropy of the smoothed tidal field. These quantities are physically connected, but they are not interchangeable: a halo can reside in an anisotropic matter distribution without its own major axis being perfectly aligned with that distribution. They should therefore be viewed as different projections of the same anisotropic large-scale-structure formation process rather than as independent environmental mechanisms.
     
     Our main goal is to study anisotropic secondary bias (ASB), defined here as the dependence of secondary bias on direction relative to the halo major axis. Unlike ordinary secondary bias (OSB), ASB probes not only whether halo properties correlate with environment, but also how this connection depends on orientation. Because ASB is a directional generalization of OSB, we first use orientation-averaged OSB as a baseline. This baseline shows which environmental quantities control the overall clustering dependence on halo properties. We then ask whether the same quantities are also connected to ASB. The comparison is designed to determine whether OSB and ASB respond to the same or to different manifestations of the anisotropic cosmic environment. In this way, OSB serves as a control experiment rather than as a separate focus.

     This approach is related to, but distinct from, previous work on anisotropic assembly bias and orientation-dependent selection effects in redshift-space clustering \citep{2009Hirata,2019Obuljen,2020Obuljen,2023Lamman,2024LammanY1}. Those studies showed that non-scalar halo or galaxy properties can introduce anisotropic clustering systematics; here we define the anisotropy with respect to each halo's own major axis, compare clustering parallel and perpendicular to that axis, and then test which environmental descriptor traces or controls the resulting halo-property dependence.
     
     We address three questions. First, which halo properties show significant ASB at fixed halo mass? Second, do halo-environment alignment, outer matter anisotropy, or tidal anisotropy explain OSB? Third, do any of these environmental manifestations trace or reduce the orientation-dependent ASB signal?
     
     This distinction matters observationally. Orientation- or property-dependent selection can contaminate redshift-space distortion measurements \citep{2009Hirata,2019Obuljen,2020Obuljen,2023Lamman,2024LammanY1}, while orientation-dependent halo-matter profiles and correlations between halo shape and large-scale structure enter weak-lensing and intrinsic-alignment modeling \citep{2006Heymans,2009Okumura,2018Osato}. ASB therefore provides a useful way to identify which halo properties are most likely to generate orientation-dependent clustering systematics.
     
     As a robustness check, we also examine halo-definition systematics. Halo boundaries and particle membership can change halo secondary properties and secondary-bias measurements (\citealt{2017Villarreal};  \citealt{2021Diemer}; \citealt{2020Fielder}; \citealt{2023Mezini};  \citealt{2022Garcia};  \citealt{2023Diemer}). Here we focus on particle membership, which is directly related to the low-mass spin-bias tension and to the UDG spin-bias interpretation of \cite{2026Ma_UDGspin}.
     
     Section \ref{section:data} describes the simulation, halo definitions, halo properties, environmental alignment and anisotropy measures, and clustering measurements. Section \ref{section:ASB} first establishes the OSB baseline and then presents the ASB measurements and environmental conditioning tests. Section \ref{section:discussion} discusses the physical interpretation, halo-definition effects, and observational implications. Section \ref{section:conclusion} summarizes our conclusions.
     
     \section{Data}
     \label{section:data}
     
     \subsection{Simulation}
     
     We use the $z=0$ TNG300-1-Dark simulation, the dark-matter-only counterpart of TNG300-1 (\citealt{2018Marinacci}; \citealt{2018Naiman}; \citealt{2018Nelson}; \citealt{2018Springel}; \citealt{2018Pillepich}). The box has side length $205h^{-1}{\rm Mpc}$ and contains $2500^3$ dark-matter particles of mass $7\times 10^7{\rm M}_{\odot}$. The adopted cosmology is $\Omega_m=0.3089$, $\Omega_b=0.0486$, $\Omega_{\Lambda}=0.6911$, $h=0.6774$, $\sigma_8=0.8159$, and $n_s=0.9667$, consistent with the Planck 2015 cosmology \citep{Planck2016CosmologicalParameters}. The public catalogues identify friends-of-friends (FoF) groups \citep{1985Davis} and gravitationally bound substructures with SUBFIND \citep{2001Springel_SUBFIND}; merger trees are taken from SUBLINK \citep{2015Rodriguez}.
     
     \subsection{Halo definitions}
     \label{data:halo_def}
     
     We use three halo definitions in the main text. The SOall definition is a spherical-overdensity host halo centred on the most-bound particle of the SUBFIND central object, using all dark-matter particles within $R_{\rm vir}$; it therefore includes unbound particles and particles in substructures. The SObound definition uses the same SO host boundary but retains only particles gravitationally bound to the host, while still keeping bound particles in substructures. The SubFind definition is the bound central subhalo returned by SUBFIND, and therefore removes satellite substructures from the particle set.
     
     Unless stated otherwise, the measurements in Section \ref{section:ASB} use SObound as the reference halo definition: halo shape, concentration, orientation, and environmental quantities are evaluated for the SObound SO hosts, and formation redshifts are taken from their merger-tree histories. Spin is the explicit exception. Following the notation of \cite{2026Ma_UDGspin}, we show both $\lambda_{\rm b}$, the SObound spin, and $\lambda_{\rm a}$, the SOall spin, to isolate the effect of unbound particles. Section \ref{sec:definition} then summarizes the halo-definition robustness test.
     
     \subsection{Halo properties}\label{data:halo_property}
     
     We use the following halo properties.
     
     \begin{itemize}
          \item {Virial mass $M_{\rm vir}$ is the mass within the virial radius for the particle set specified by the halo definition, using the virial overdensity threshold of \cite{1998Bryan}. The virial velocity is $v_{\rm vir}=\sqrt{GM_{\rm vir}/R_{\rm vir}}$.}
          \item {Formation redshift $z_{1/2}$ is the redshift at which the halo first reaches half of its final mass, estimated by linearly interpolating between the two adjacent merger-tree snapshots \citep{2005Gao}.}
          \item {Concentration is measured with $c_{vmax}$, which is more robust than direct NFW fitting for some particle sets \citep{2012Prada,2011Klypin}: 
             \begin{equation}
                 \frac{c_{vmax}}{f(c_{vmax})} = \frac{v_{\max}^2}{v_{\rm vir}^2} \frac{r_{\max}/r_s}{f(r_{\max}/r_s)},
             \end{equation}
             \begin{equation}
                 f(x) \equiv \ln (1+x)-\frac{x}{1+x},
             \end{equation}
              where $r_{\max}=2.1626r_s$ for an NFW profile \citep{navarroStructureColdDark1996,navarroUniversalDensityProfile1997}. Direct NFW concentrations give consistent results.
          }
          \item {Spin is the dimensionless angular-momentum parameter of \cite{2001Bullock},
             \begin{equation}
           \lambda=\frac{|J|}{\sqrt{2} M_{\rm vir} V_{\rm vir} R_{\rm vir}},
           \end{equation}
           where $|J|$ is measured from the chosen particle set within $R_{\rm vir}$. For the SO definitions introduced in Section \ref{data:halo_def}, $\lambda_{\rm a}$ is computed with the SOall particle set and $\lambda_{\rm b}$ with the SObound particle set. The same Bullock formula is used for the other halo definitions. We use the core velocity as the halo velocity following \cite{2013Behroozi}; using the bulk velocity changes the spin values but not the spin-bias trends.
          }
          \item {Shape is estimated iteratively from the weighted inertia tensor \citep{2006Allgood},
              \begin{equation}
                \tilde{I}_{i j} \equiv \sum^{N}_k \frac{x_{i, k} x_{j, k}}{r_k^2},
                \label{eq:inertia_tensor}
            \end{equation}
            where $x_{i,k}$ is the position of the $k$th particle relative to the halo centre. The ellipsoidal axes satisfy $a \geq b \geq c$, with $q=b/a$ and $s=c/a$, and are obtained from the tensor eigenvalues. The weight is the elliptical distance
            \begin{equation}
                r_k=\sqrt{x_k^2+y_k^2 / q^2+z_k^2 / s^2}.
            \end{equation}
            We use particles within $R_{\rm vir}$ and define triaxiality as $T=(1-q^2)/(1-s^2)$.
         }
     \end{itemize}
     
     %Due to the difficulty of recalculating the merger tree, the formation time is obtained by the merger tree based on the SUBLINK \citep{2015Rodriguez}. The other properties are estimated by different particle sets within the virial radius.
     
     \subsection{Environmental alignment and anisotropy measures}
     \label{data:anisotropy}
     
     The deformation tensor is the Hessian of the rescaled gravitational potential $\phi$ \citep{2007HahnB,2009Forero_cosmicweb},
     \begin{equation}
         T_{\alpha \beta}=\frac{\partial^2 \phi}{\partial r_\alpha r_\beta},
     \end{equation}
     where $\phi$ is rescaled by $4\pi G\bar{\rho}=3\Omega_mH_0^2/2$. With this normalization, the Poisson equation becomes $\nabla^2\phi=\delta$. The density field is smoothed with a halo-dependent $4R_{\rm vir}$ Gaussian kernel so that the tidal field traces the large-scale environment on a scale tied to the size of each halo. We order the eigenvalues as $\lambda_1\geq \lambda_2\geq \lambda_3$, so that the eigenvector $\mathbf{e}_3$ corresponds to the slowest-collapse direction.
     
     We use five environmental descriptors, grouped into three physical manifestations. All comparisons involving scalar quantities derived from these descriptors are made at fixed halo mass; below they are used either as secondary descriptors or as conditioning variables.
     
     \begin{itemize}
         \item {Halo-environment alignment describes the part of the environment that is aligned with the halo major axis. We use two measures. The density ratio $n_{\parallel}/n_{\perp}$ compares matter along and away from the halo major axis: $n_{\parallel}$ is estimated from matter in the radial range $r=5\text{--}12\,h^{-1}{\rm Mpc}$ and within $\theta<45^{\circ}$ of the halo major axis, while $n_{\perp}$ uses matter in the same radial range but within $\theta>45^{\circ}$. Values $n_{\parallel}/n_{\perp}>1$ indicate that the surrounding matter distribution is preferentially extended along the halo major axis. We also use the tidal-field alignment $A\cdot e_3\equiv |\mathbf{A}\cdot\mathbf{e}_3|$, where $\mathbf{A}$ is the halo major-axis unit vector and the absolute value is used because the halo major axis is unoriented. For haloes in filamentary environments, larger $A\cdot e_3$ indicates stronger alignment with the local filament direction. Both quantities are therefore halo-axis alignment descriptors: they are sensitive to the component of the surrounding structure aligned with the halo, but they do not by themselves measure the full anisotropy amplitude of the outer matter distribution.}
          \item {Outer matter anisotropy describes the shape of the matter distribution outside the virialized halo. We measure the outer matter major axis $\mathbf{A}_{4R_{\rm vir}}$ and outer matter axis ratio $c/a(4R_{\rm vir})$ from dark matter in the shell $2R_{\rm vir}<r<4R_{\rm vir}$. Operationally, we evaluate the weighted inertia tensor of Eq.~(\ref{eq:inertia_tensor}) for the particles in this shell without iterative re-scaling, and define $\mathbf{A}_{4R_{\rm vir}}$ and $c/a(4R_{\rm vir})$ from its eigenvectors and eigenvalues. The quantity $|\mathbf{A}_{4R_{\rm vir}}\cdot\mathbf{e}_3|$ tests how well the outer matter major axis follows the large-scale tidal direction, while smaller $c/a(4R_{\rm vir})$ indicates a more anisotropic outer matter distribution.}
          \item {Tidal anisotropy describes the anisotropy of the smoothed tidal field. The local tidal anisotropy is defined from the same deformation tensor \citep{2018Paranjape_tidalaniso} as
               \begin{equation}
                   \alpha=\sqrt{q^2} /(1+\delta),
               \end{equation}
               where $q^2=\frac{1}{2}[(\lambda_1-\lambda_2)^2+(\lambda_2-\lambda_3)^2+(\lambda_1-\lambda_3)^2]$ is the tidal shear and $\delta=\lambda_1+\lambda_2+\lambda_3$ is the overdensity.
          }
     \end{itemize}
     
     \subsection{Halo statistics}
     
     Secondary bias is measured with the relative bias \citep{2018Salcedo},
     
     \begin{equation}
     b^{rel}_{S}\left(r|\mathrm{M}_{\mathrm{vir}}\right)=\sqrt{\frac{\xi_{hh}\left(r|\mathrm{M}_{\mathrm{vir}},\mathrm{S}\right)}{\xi_{hh}\left(r|\mathrm{M}_{\mathrm{vir}}\right)}},
     \end{equation}
     where $\xi_{hh}(r|M_{\rm vir},S)$ is the halo autocorrelation of a secondary-property-selected subsample at fixed mass. Unless otherwise stated, secondary-property-selected samples are the upper and lower quartiles of each halo property in a narrow mass bin. Errors combine 50 bootstrap resamplings with Poisson errors. All clustering amplitudes are averaged over $5$--$12h^{-1}{\rm Mpc}$. We use haloes with $10^{10.7}<M_{\rm vir}/(h^{-1}{\rm M}_{\odot})<10^{14.0}$; the lower limit keeps at least 1000 particles per halo, and spin-bias tests are insensitive to the particle-number noise estimated by \cite{2017Benson}.
     
     To measure orientation dependence, we use the alignment correlation function (ACF), which adds the angle $\theta$ between the halo major axis and the pair-separation vector to the usual two-point correlation function \citep{2008Paz,2009Faltenbacher,2011Paz}:
     \begin{equation}
     \xi\left(r|\mathrm{M}_{\mathrm{vir}},S, \theta\right)=\frac{N_{\mathcal{R}}}{N_R} \frac{Q R\left(r|\mathrm{M}_{\mathrm{vir}},S, \theta\right)}{Q \mathcal{R}\left(r|\mathrm{M}_{\mathrm{vir}},S, \theta\right)}-1,
        \label{eq:xi_aniso}
     \end{equation}
     where $Q$ is the selected halo sample, $R$ is the reference halo sample, $\mathcal{R}$ is the random catalogue, and $N_R$ and $N_{\mathcal{R}}$ are the numbers of objects in the reference and random samples. We set $Q=R$ for the halo autocorrelation in each mass, property, and orientation bin. The same angular selections are applied to the full sample and to every secondary-property subsample: pairs with $\theta<45^{\circ}$ are denoted parallel to the halo major axis, while pairs with $\theta>45^{\circ}$ are denoted perpendicular.
     
     We use two normalized quantities to separate OSB from orientation dependence:
     \begin{equation}
     b^{S,\theta}_{S}\left(r|\mathrm{M}_{\mathrm{vir}},S,\theta\right)=\sqrt{\frac{\xi_{hh}\left(r|\mathrm{M}_{\mathrm{vir}},S,\theta\right)}{\xi_{hh}\left(r|\mathrm{M}_{\mathrm{vir}},S\right)}},
     \end{equation}
     \begin{equation}
     b^{S,\theta}_{\theta}\left(r|\mathrm{M}_{\mathrm{vir}},S,\theta\right)=\sqrt{\frac{\xi_{hh}\left(r|\mathrm{M}_{\mathrm{vir}},S,\theta\right)}{\xi_{hh}\left(r|\mathrm{M}_{\mathrm{vir}},\theta\right)}}.
     \end{equation}
     Here $b^{S,\theta}_{S}$ tests whether the orientation-dependent clustering depends on the secondary property, while $b^{S,\theta}_{\theta}$ tests whether secondary bias depends on direction after normalizing by the overall anisotropic clustering. By definition, $b^{S,\theta}_{\theta}=1$ means no secondary bias in that direction. We refer to both quantities as anisotropic secondary bias (ASB).
     
     \section{Results}
     \label{section:ASB}
     
     \subsection{OSB as an environmental baseline}
     
     Unless explicitly stated otherwise, the measurements in this section use the SObound definition introduced in Section \ref{data:halo_def}; the $\lambda_{\rm a}$ panel is the SOall spin included for comparison.
     
     We first use OSB as a baseline for the environmental tests. The goal is to determine which of the three environmental manifestations in Section \ref{data:anisotropy} controls the overall clustering dependence on halo properties before asking whether the same quantities are also connected to, or can reduce, ASB. This subsection is therefore a control step: OSB identifies which environmental variable affects the orientation-averaged clustering amplitude, while the following subsections test whether the same variable also reduces or leaves intact the directional signal. For a controlled quantity $\Xi$, the conditioned secondary bias is
     \begin{equation}
     b^{rel}_{S}|\Xi=\sqrt{\frac{\xi_{hh}\left(\mathrm{M}_{\mathrm{vir}},\mathrm{S}|\Xi\right)}{\xi_{hh}\left(\mathrm{M}_{\mathrm{vir}}\right|\Xi)}},
     \end{equation}
     where $\Xi$ is the controlled quantity. At fixed mass, the $\Xi$ distribution is matched between the upper quartile, lower quartile, and full samples. Operationally, the matching is performed independently in each halo-mass bin. We divide $\Xi$ into 10 bins and construct matched comparison samples so that the $\Xi$ distribution of each property-selected subsample follows that of the corresponding full mass-bin sample before recomputing the correlation functions and bootstrap uncertainties. The matching is applied before any pair counts are accumulated, so the numerator and denominator of the conditioned statistic are measured from catalogues with the same halo-mass selection and the same controlled-quantity distribution. Bootstrap resampling is then repeated for the matched catalogues, and the quoted error bars therefore correspond to the conditioned measurements. The same matching prescription is used for all environmental descriptors, allowing the strength of the residual OSB to be compared across halo-environment alignment, outer matter anisotropy, and tidal anisotropy.
     
     \begin{figure*}[!t]
          \includegraphics[width=1.0\textwidth]{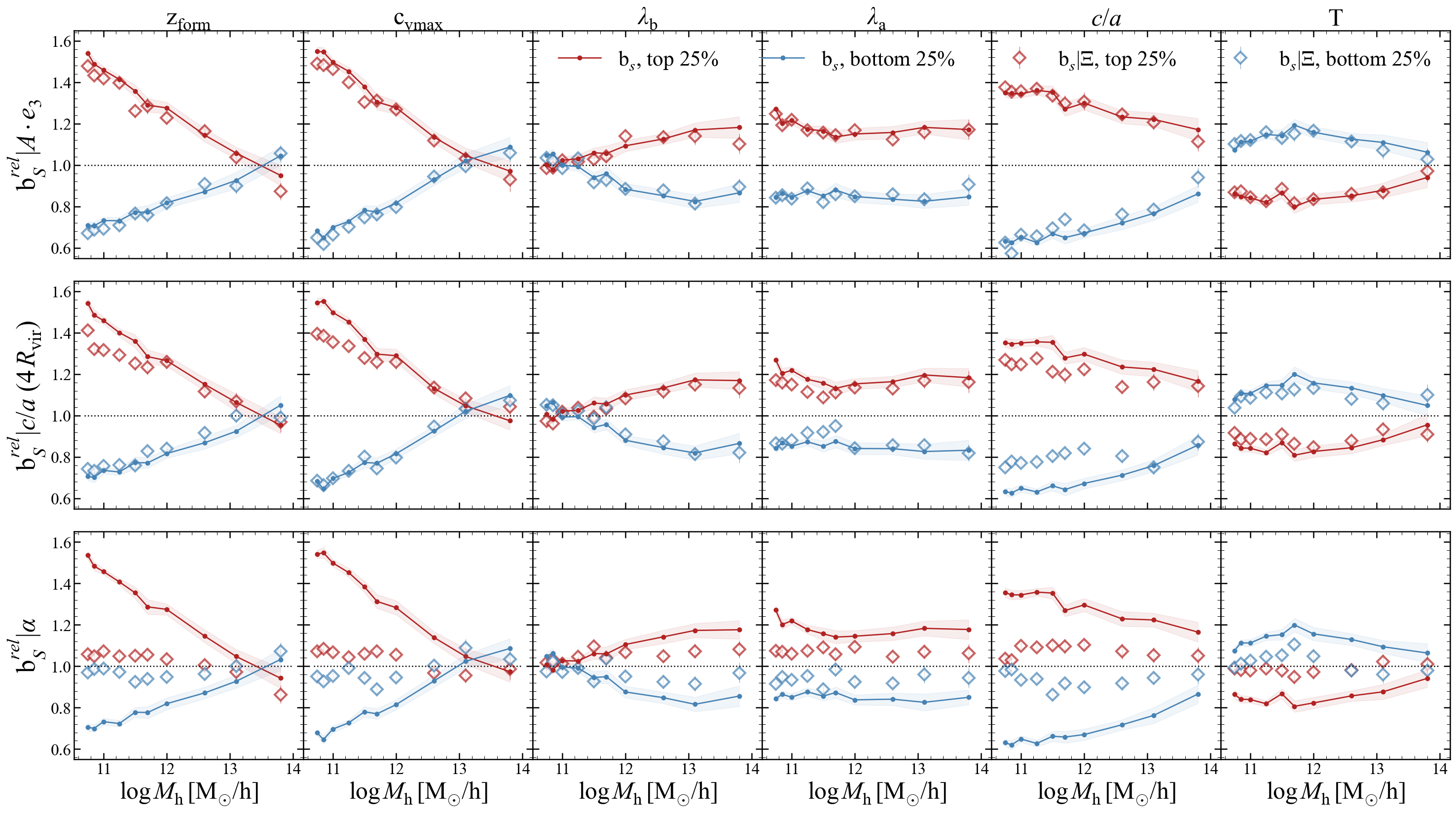}
         \caption{Three environmental controls of OSB. Rows show secondary bias before and after controlling for halo-environment alignment $A\cdot e_3$ (upper), outer matter anisotropy $c/a(4R_{\rm vir})$ (middle), and tidal anisotropy $\alpha$ (lower). Columns, from left to right, show $z_{\rm form}$, $c_{vmax}$, $\lambda_{\rm b}$, $\lambda_{\rm a}$, $c/a$, and $T$. Red and blue denote the upper and lower 25\% subsamples of each halo property, respectively. Solid curves with filled circles show the original relative bias $b^{rel}_{S}$, while open diamonds show the conditioned measurement after matching the row-specific environmental descriptor between each property-selected subsample and the full halo sample at fixed mass. The horizontal dotted line marks $b^{rel}_{S}=1$, i.e. no secondary bias. Shaded bands and error bars show bootstrap uncertainties.}
         \label{fig:matteraniso_all}
     \end{figure*}
     
     Figure \ref{fig:matteraniso_all} summarizes the OSB baseline and the three environmental controls. The solid curves, repeated in each row, show the original OSB measurements for six halo properties. At low mass, early-forming, high-concentration, low-spin, more elongated, and more triaxial haloes are generally more strongly clustered than their counterparts; the concentration and formation-time trends reverse toward high mass, while the spin and shape trends persist. The two spin definitions differ in the way already found by \cite{2026Ma_UDGspin}: the low-mass spin-bias inversion appears for the bound-particle spin $\lambda_{\rm b}$ but not for the all-particle spin $\lambda_{\rm a}$.

     The upper row of Figure \ref{fig:matteraniso_all} shows the representative result after matching the halo-environment alignment $A\cdot e_3$. The conditioned points closely follow the original OSB curves over most mass bins, and the changes are generally smaller than the separation between the upper- and lower-quartile samples. Matching $n_{\parallel}/n_{\perp}$ gives the same qualitative result. Halo-environment alignment is therefore correlated with some halo properties, as shown below, but it does not account for OSB. We return to the relation between the two halo-alignment descriptors in Section \ref{sec:asb_env}.
     
    The outer matter anisotropy tests give a similar null result for OSB. The middle row of Figure \ref{fig:matteraniso_all} shows that matching $c/a(4R_{\rm vir})$, the anisotropy amplitude of the outer matter distribution, leaves the OSB curves close to their original values. The $|\mathbf{A}_{4R_{\rm vir}}\cdot\mathbf{e}_3|$ control gives the same qualitative result. Thus neither halo-axis alignment nor outer matter anisotropy is the dominant driver of OSB.
     
     Tidal anisotropy behaves differently. The bottom row of Figure \ref{fig:matteraniso_all} shows that matching $\alpha$ substantially suppresses OSB for all six halo properties, bringing the upper- and lower-quartile samples much closer to unity over most of the mass range. This suppression is not confined to a particular halo-mass scale, although residual bias remains, especially for spin and halo shape. We therefore recover the expected result that tidal anisotropy captures an important part, but not all, of OSB.
     
     This baseline will be used below to interpret ASB. If ASB were simply the directional manifestation of the same environmental physics that controls OSB, then matching the relevant environmental quantity should also erase the orientation-dependent signal.
     
     \subsection{\texorpdfstring{ASB measurements}{ASB measurements}}
     
     \begin{figure*}[!t]
          \includegraphics[width=1.0\textwidth]{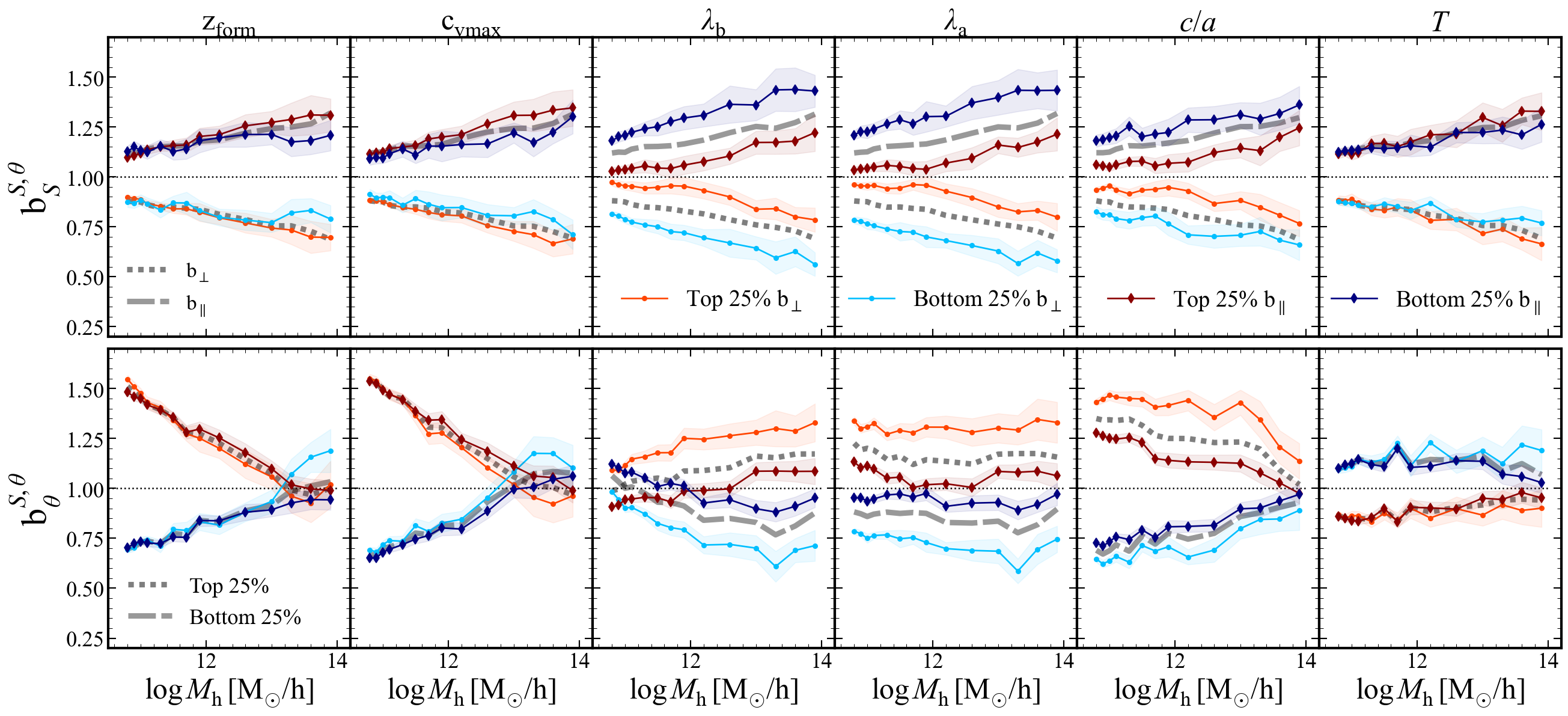}
         \caption{Anisotropic secondary bias as a function of halo mass. Columns, from left to right, show formation redshift $z_{\rm form}$, concentration $c_{vmax}$, bound-particle spin $\lambda_{\rm b}$, all-particle spin $\lambda_{\rm a}$, minor-to-major axis ratio $c/a$, and triaxiality $T$. Upper panels show $b^{S,\theta}_{S}$, the orientation-dependent clustering normalized by the secondary-property-selected sample; lower panels show $b^{S,\theta}_{\theta}$, the secondary-bias signal measured separately parallel ($\theta<45^{\circ}$) and perpendicular ($\theta>45^{\circ}$) to the halo major axis. For the colored curves, dark red and pink show the upper 25\% subsample measured in the parallel and perpendicular directions, while dark blue and cyan show the corresponding lower 25\% subsample. In the upper panels, gray dashed and dotted curves show the full-sample alignment signals $b_{\parallel}$ and $b_{\perp}$, respectively. In the lower panels, gray dotted and dashed curves show the direction-averaged relative bias $b^{rel}_{S}$ for the upper and lower 25\% subsamples, respectively.}
         \label{fig:secb_theta_all}
     \end{figure*}
     
     Figure \ref{fig:secb_theta_all} summarizes the ASB measurements. The gray curves in the upper panels show the overall alignment signal, which strengthens with halo mass as expected from previous alignment studies \citep{2009Faltenbacher}. Formation time, concentration, and triaxiality show weak or no ASB. Formation time has only a small high-mass trend, and concentration shows at most a mild orientation dependence.
     
     The two spin definitions, $\lambda_{\rm b}$ and $\lambda_{\rm a}$, and axis ratio show the clearest ASB. The spin panels first reproduce the definition dependence found by \cite{2026Ma_UDGspin}: the bound-particle spin $\lambda_{\rm b}$ shows a low-mass spin-bias inversion, whereas the all-particle spin $\lambda_{\rm a}$ does not. Figure \ref{fig:secb_theta_all} adds the new result that the $\lambda_{\rm b}$ inversion is anisotropic. It appears only for pairs parallel to the halo major axis, where low-$\lambda_{\rm b}$ haloes are more strongly clustered than high-$\lambda_{\rm b}$ haloes; the perpendicular measurement does not show the same inversion. Thus the bound-particle spin-bias inversion is tied to the orientation dependence of halo clustering. More elongated haloes are also more strongly aligned than rounder haloes, consistent with previous work on shape-dependent alignment \citep{2009Okumura,2012Schneider,2021Akitsu}. By contrast, triaxiality does not produce a clear orientation-dependent signal.
     
     Thus ASB is mainly associated with spin and halo elongation rather than formation history or concentration. This separation is consistent with the picture that mass history and virialization are more closely tied to local density, while shape, spin, and orientation retain stronger sensitivity to the large-scale environment \citep{2012Codis,2018Ganeshaiah,2024Storck,2021Chen}; see also \citealt{2021Ganeshaiah_ZOMGIII}. These measurements identify which halo properties display ASB; we next compare the environmental manifestations that may be connected to these signals.
     
     \subsection{\texorpdfstring{Environmental connections of ASB}{Environmental connections of ASB}}
     \label{sec:asb_env}
     
     We now examine ASB in relation to the three environmental manifestations introduced in Section \ref{data:anisotropy}. Halo-environment alignment is considered first because it is geometrically closest to the ASB measurement, but it is treated as one member of the same environmental comparison rather than as a separate ingredient.
     
     \begin{figure*}[!t]
      \includegraphics[width=0.95\textwidth]{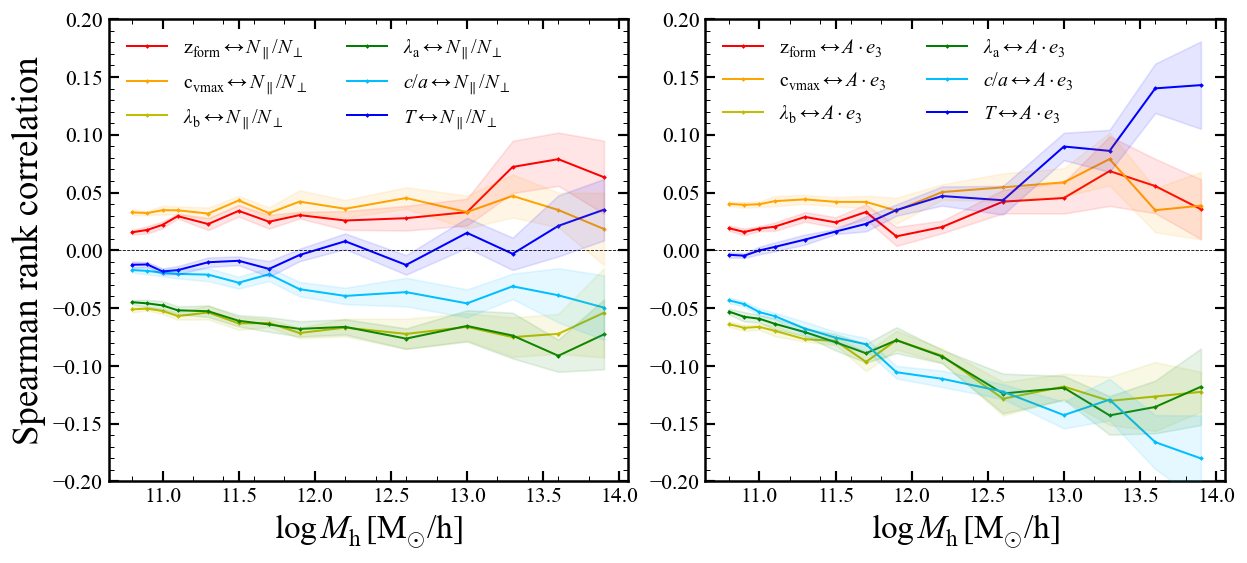}
 \caption{Spearman rank correlations between halo secondary properties and halo-environment alignment measures as a function of halo mass. The left panel uses the density-alignment ratio $n_{\parallel}/n_{\perp}$, and the right panel uses the absolute alignment $A\cdot e_3\equiv |\mathbf{A}\cdot\mathbf{e}_3|$ between the halo major axis and the slowest-collapse direction of the tidal field. In both panels, red, orange, olive, green, cyan, and blue curves show correlations for $z_{\rm form}$, $c_{vmax}$, $\lambda_{\rm b}$, $\lambda_{\rm a}$, $c/a$, and $T$, respectively. The horizontal dotted line marks zero correlation; positive values mean that the halo property increases with the alignment measure, and negative values mean the opposite. Shaded regions indicate bootstrap uncertainties.}
     \label{fig:AAB_spearman}
     \end{figure*}
     
     Figure \ref{fig:AAB_spearman} quantifies the connection between ASB and halo-environment alignment at the halo-by-halo level. In each mass bin, it shows the Spearman rank correlation between each secondary property and the two halo-environment alignment descriptors defined in Section \ref{data:anisotropy}: the density-alignment ratio $n_{\parallel}/n_{\perp}$ and the tidal-field alignment $A\cdot e_3$. The two panels show similar trends, with $A\cdot e_3$ giving slightly stronger correlations for most of the properties most relevant to ASB. Spin and minor-to-major axis ratio have the strongest negative correlations with both alignment measures, meaning that lower-spin and more elongated haloes tend to have stronger environmental alignment with their major axes. Formation time and concentration correlate only weakly, while triaxiality shows a positive correlation that becomes more visible at high mass. These trends help identify why spin and shape display the strongest ASB, even though halo-environment alignment does not explain OSB.
     
     We next apply matched-control tests to all three environmental manifestations, halo-environment alignment, outer matter anisotropy and tidal anisotropy. For a controlled quantity $\Xi$, the conditioned ASB is
     \begin{equation}
     b^{S,\theta}_{S}|\Xi=\sqrt{\frac{\xi_{hh}\left(\mathrm{M}_{\mathrm{vir}},\mathrm{S}, \theta|\Xi\right)}{\xi_{hh}\left(\mathrm{M}_{\mathrm{vir}},S\right|\Xi)}}.
     \end{equation}
     Here the conditioning is applied before the angular correlation functions are measured: within each mass bin and secondary-property subsample, the parallel and perpendicular pair counts are recomputed using samples matched in $\Xi$ in the same way as described above. Thus the environmental control is imposed at the halo-catalogue level before the same matched halo catalogues are used to form the parallel and perpendicular pair counts. This ensures that any residual separation between the parallel and perpendicular curves reflects orientation dependence at fixed mass, fixed secondary-property selection, and matched environmental descriptor.
     
     \begin{figure*}[!t]
          \includegraphics[width=1.0\textwidth]{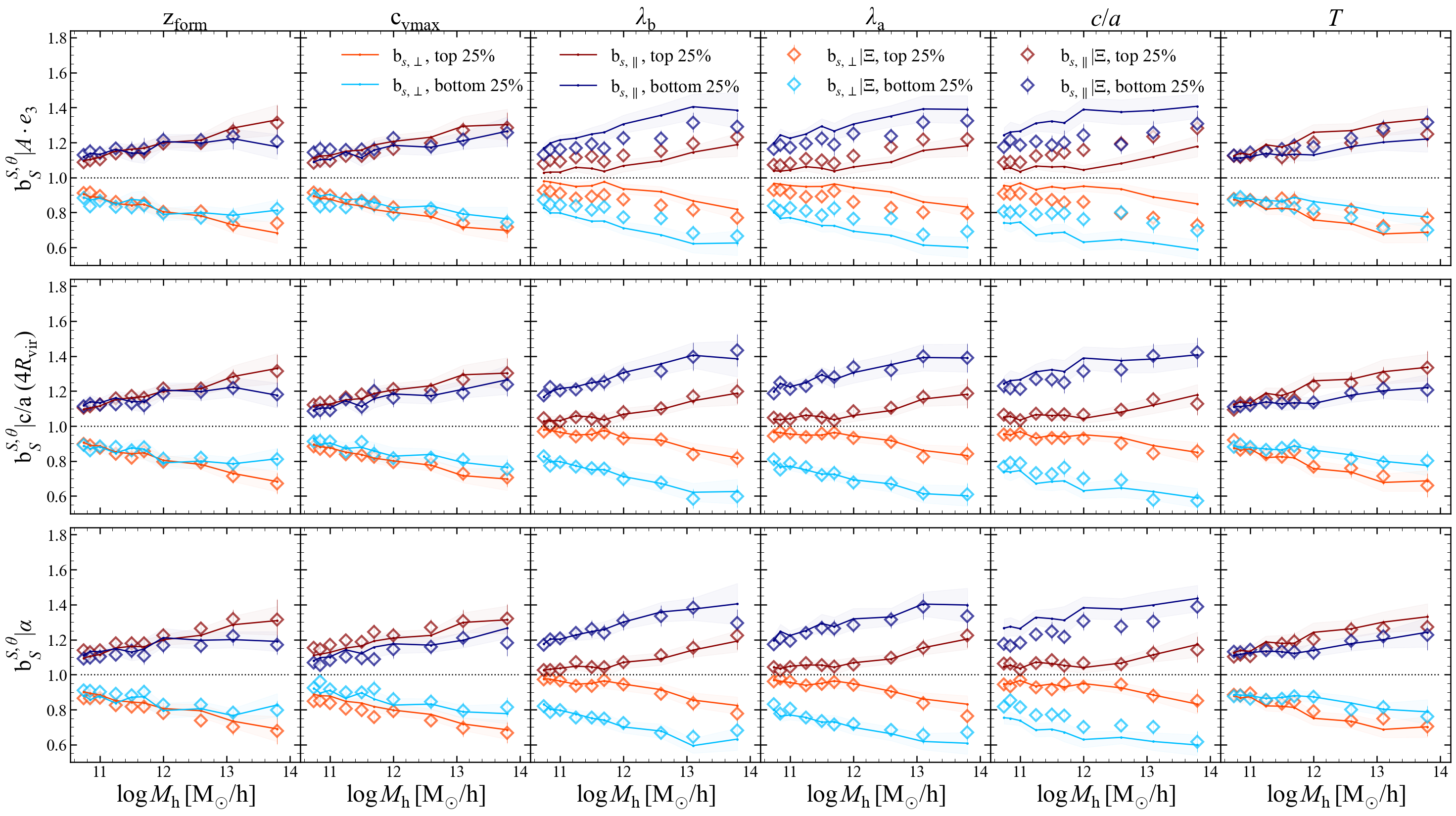}
         \caption{Three environmental controls of ASB. Rows show $b^{S,\theta}_{S}$ before and after controlling for halo-environment alignment $A\cdot e_3$ (top), outer matter anisotropy $c/a(4R_{\rm vir})$ (middle), and tidal anisotropy $\alpha$ (bottom). Columns, from left to right, show $z_{\rm form}$, $c_{vmax}$, $\lambda_{\rm b}$, $\lambda_{\rm a}$, $c/a$, and $T$. Solid curves show the original ASB measurements, and open diamonds show the corresponding measurements after matching the row-specific environmental descriptor at fixed mass. Dark red and dark blue show the upper and lower 25\% subsamples measured parallel to the halo major axis, while pink and cyan show the corresponding upper and lower 25\% subsamples measured perpendicular to the major axis. The horizontal dotted line marks $b^{S,\theta}_{S}=1$. Shaded bands and error bars show bootstrap uncertainties.}
         \label{fig:secb_theta_sa_all}
     \end{figure*}
     
     Figure \ref{fig:secb_theta_sa_all} shows the results of these matched-control tests. The top row shows that matching halo-environment alignment $A\cdot e_3$ substantially reduces the ASB signals for spin and halo elongation, although residual orientation dependence remains in some mass bins. This indicates that ASB is most directly traced by the coupling between halo properties and the part of the environment aligned with the halo major axis.
     
     The middle row of Figure \ref{fig:secb_theta_sa_all} shows that matching $c/a(4R_{\rm vir})$ leaves the ASB signal largely intact. The conditioned measurements retain the same separation between parallel and perpendicular directions for the spin- and shape-selected samples, with changes much smaller than the original ASB amplitude. The outer-matter/tidal-alignment ASB control gives the same conclusion. Thus ASB is not simply the imprint of the anisotropy amplitude or tidal alignment of the outer matter distribution.
     
     The bottom row of Figure \ref{fig:secb_theta_sa_all} shows that matching $\alpha$ leaves the ASB signal largely intact: the conditioned measurements retain the same direction dependence and remain well separated for the spin- and shape-selected samples. The conditioning changes some amplitudes slightly, but it does not remove the orientation-dependent signal. Thus tidal anisotropy affects OSB much more strongly than it affects the relation between halo orientation and secondary halo properties.
     
      In summary, the three environmental manifestations connect to ASB in different ways. Halo-environment alignment is closely related to the halo properties that show ASB, especially spin and elongation, and absorbs much of the corresponding ASB signal, but it does not explain OSB. Outer matter anisotropy, measured independently of the halo major axis, does not erase ASB. Tidal anisotropy explains a substantial part of OSB but has little effect on ASB. A possible physical interpretation is discussed in Section \ref{sec:tidalani}.
     
     \section{Discussion}
     \label{section:discussion}
     
     \subsection{\texorpdfstring{Environmental origin of OSB and ASB}{Environmental origin of OSB and ASB}}
     \label{sec:tidalani}
     
     The results above should be interpreted as different manifestations of one anisotropic large-scale-structure formation process, rather than as evidence for unrelated physical origins of OSB and ASB. OSB describes how secondary halo properties modulate the orientation-averaged clustering amplitude at fixed mass. ASB asks how the same property dependence changes with direction relative to the halo major axis. The comparison is therefore useful precisely because it tests whether the environmental descriptor that controls the scalar, orientation-averaged signal also controls the directional signal.
     
     Tidal fields are a leading explanation for OSB \citep{2007Wang,2009Hahn_tidalforce,2011Wang}. They can arise from massive neighbours, including splashback-related environments \citep{2018Salcedo,2021Tucci}, or from the cosmic web, where stalled haloes in thick filaments accrete differently from more isolated systems \citep{2017Borzyszkowski,2018Musso_EPScosmicweb_AB,2018Paranjape_tidalaniso,2019Ramakrishnan_tidalaniso}. Our baseline tests support this picture: matching tidal anisotropy $\alpha$ suppresses much of OSB.
     
     ASB does not follow the same response. If ASB were merely the directional expression of the tidal-anisotropy dependence that produces OSB, then matching $\alpha$ should also erase the orientation-dependent signal. Instead, spin- and shape-dependent ASB remain nearly unchanged. Thus tidal anisotropy appears to regulate part of the overall clustering amplitude, while ASB is more directly tied to the coupling between halo orientation and secondary halo properties. This contrast does not imply that tidal anisotropy and halo alignment are independent physical events; rather, it shows that the scalar tidal-anisotropy measure $\alpha$ captures the part of the anisotropic environment most relevant for OSB, but not the projection most relevant for ASB.
     
     Halo-environment alignment plays a different role. The quantities $A\cdot e_3$ and $n_{\parallel}/n_{\perp}$ are closely connected to spin and shape, and this connection helps identify why spin- and shape-selected haloes show the strongest ASB. However, these quantities do not erase OSB when used as conditioning variables. When used as ASB controls, by contrast, matching $A\cdot e_3$ absorbs much of the spin- and shape-dependent orientation signal. They should therefore not be interpreted as general measures of matter anisotropy; they describe the component of the environment aligned with the halo major axis. In this sense, they are the environmental manifestation most directly matched to the geometry of the ASB measurement itself.
     
     The outer matter analysis separates this alignment effect from the anisotropy amplitude of the surrounding matter. The axis ratio $c/a(4R_{\rm vir})$, measured in the shell $2R_{\rm vir}<r<4R_{\rm vir}$, provides a direct measure of how anisotropic the outer matter distribution is, independent of whether its major axis is aligned with the halo major axis. Matching this quantity leaves ASB nearly unchanged. This result suggests that ASB is not simply caused by haloes living in more anisotropic matter environments; rather, it reflects how halo spin and shape are coupled to halo orientation within the cosmic web. More generally, these conditioning tests should be interpreted as single-descriptor tests rather than complete causal interventions. If matching one scalar descriptor leaves ASB unchanged, the result means that this particular projection of the environment, measured with the adopted scale and aperture, is not sufficient to absorb the orientation-dependent signal; other correlated or multivariate environmental information may still contribute.
     
     Several limitations should be kept in mind. Our measurements use one $z=0$ dark-matter-only simulation and therefore do not test redshift evolution, baryonic modifications of halo shape and spin, or possible volume dependence at the highest masses. The highest-mass bins are also the most susceptible to sample variance, so the qualitative comparison between unconditioned and conditioned measurements is more robust than the precise amplitude in any single high-mass bin. The environmental descriptors also depend on methodological choices such as the tidal-field smoothing scale, the outer-matter aperture, the angular cuts used to define halo-axis alignment, and the number of bins used in the matching procedure. The results should be interpreted as halo-level trends in a $\Lambda$CDM simulation rather than as a direct prediction for any particular galaxy sample. Hydrodynamic simulations and observationally selected mock catalogues will be needed to determine how much of the halo-level ASB propagates to galaxies.
     
     \subsection{Halo-definition systematics}
     \label{sec:definition}
     
     We use SObound as the reference definition throughout most of the analysis, while keeping the all-particle spin $\lambda_{\rm a}$ as a direct comparison to the bound-particle spin $\lambda_{\rm b}$. To check halo-definition systematics, we repeated the main measurements with the SOall, SObound, and SubFind definitions introduced in Section \ref{data:halo_def}. Appendix \ref{app:halo_def} shows one OSB comparison among SOall, SObound, and SubFind, and the direct ASB comparison between SOall and SObound for concentration, spin, shape, and triaxiality. This comparison is intended only as a robustness test for the ASB results, because the detailed effect of particle membership and halo boundary on spin OSB has already been studied in \cite{2026Ma_UDGspin}.
     
     The strongest halo-definition dependence is seen in spin OSB. Including unbound particles raises the all-particle spin of some haloes in dense environments and changes the familiar low-mass spin-bias inversion, consistent with \cite{2026Ma_UDGspin}. Concentration shows a smaller low-mass definition dependence, while shape and triaxiality are nearly unchanged. In contrast, the ASB measurements are nearly unchanged between SOall and SObound, and using SubFind does not alter the qualitative conclusion that spin and halo elongation show the strongest orientation-dependent signals. The most visible definition-dependent change in the ASB comparison is a modest amplitude shift in the spin panel, but the ordering of the parallel and perpendicular curves and the identification of spin and elongation as the strongest ASB tracers remain unchanged. Halo definition is therefore an important caveat for interpreting spin bias and comparing halo catalogues, but it has only a minor impact on the main ASB conclusions of this paper.
     
     \subsection{Insight into the galaxy and halo connection}
     
     Galaxy-halo models that depend only on halo mass are known to leave secondary-bias systematics \citep{2007Croton,2014Zentner,2018McEwen,2018Zehavi}, motivating models with additional halo properties \citep{2013Hearin,2015Rodriguez,2015Paranjape,2016Hearin,2016Zu,2017Lehmann,2021Xu}. Recent group-catalog and DESI measurements are beginning to test these effects observationally \citep{2025Shao,2026Rodriguez}. Our results highlight two cautions for such models. First, if galaxy properties are linked to spin or concentration, the inferred secondary bias can depend on halo definition. The UDG result of \cite{2026Ma_UDGspin}, where the observed clustering is naturally reproduced when galaxy surface density is linked to an all-particle spin definition, is a concrete example, and direct observational probes of halo spin bias are also being developed \citep{2025Kim}.
     
     Second, if galaxy properties are linked to halo spin or shape, the model can inherit an orientation-dependent clustering signal. This is relevant for observed correlations between galaxy properties and large-scale alignment \citep{2009Faltenbacher,2019ChenY}, as well as for satellite anisotropic quenching studies \citep{2006Yang,2013Zhang,2021Mart,2022HadzhiyskaA}. A key question for future work is how much of the galaxy-level anisotropy follows halo assembly and orientation, and how much is reshaped by baryonic physics \citep{2023Karp}.
     
     \subsection{The effect of ASB in weak lensing and redshift-space distortion}
     
     ASB can enter two common anisotropic clustering probes. For weak lensing, the relevant issues are intrinsic alignment and orientation-dependent halo-matter profiles. \cite{2009Okumura} showed that the correlation between halo shape and orientation can change the predicted GI signal by about 15\% if neglected, and \cite{2018Osato} found that halo orientation can shift the large-scale two-halo surface-density amplitude. Our measurements support this concern for shape-selected samples and suggest that spin-selected samples may also carry orientation-dependent systematics.
     
     For redshift-space distortions (RSD), the concern is that orientation-dependent galaxy selection can mimic or contaminate the anisotropic clustering normally attributed to velocities. \cite{2009Hirata} pointed out this possibility, and later work found that ASB can be comparable to the standard RSD signal for group-selected samples \citep{2019Obuljen,2020Obuljen}, while DESI estimates indicate a smaller but redshift-dependent correction for LRG samples \citep{2024LammanY1}. Some simulations find small effects on inferred cosmological parameters \citep{2018Pillepich,2019McCarthy}. Our results point to a simple mitigation: selections tied to spin and shape should be modeled explicitly, because these properties carry the strongest ASB signals.
     
     \section{Conclusions}
     \label{section:conclusion}
     
     We have studied anisotropic secondary bias (ASB) in the $z=0$ TNG300-1-Dark simulation. ASB is defined as the dependence of secondary bias on direction relative to the halo major axis, or equivalently as the dependence of anisotropic clustering on secondary halo properties at fixed mass. We first used OSB as a baseline to test the manifestations of halo-environment alignment, outer matter anisotropy, and tidal anisotropy. We then measured ASB for formation time, concentration, two spin definitions, axis ratio, and triaxiality; tested whether the same environmental quantities trace or reduce the orientation-dependent signal; and examined halo-definition dependence as a robustness check.
     
     Our main conclusions are as follows.
     
     \begin{itemize}
          \item ASB is weak or absent for formation time, concentration, and triaxiality, but is significant for both spin definitions and minor-to-major axis ratio. Slowly rotating haloes and more elongated haloes are more strongly aligned with the large-scale filamentary structure.
          
          \item Halo-environment alignment is connected to the halo properties that show ASB but does not explain OSB. The two alignment measures considered here, $A\cdot e_3$ and $n_{\parallel}/n_{\perp}$, correlate most strongly with spin- and shape-related quantities. However, matching these quantities across subsamples leaves the OSB signal nearly unchanged, while matching $A\cdot e_3$ substantially reduces the spin- and shape-dependent ASB signals.
          
          \item The anisotropy amplitude of the outer matter distribution explains neither OSB nor ASB. Matching the outer matter axis ratio $c/a(4R_{\rm vir})$, measured in the shell $2R_{\rm vir}<r<4R_{\rm vir}$, leaves the spin- and shape-dependent ASB signals largely intact. Thus ASB is not simply caused by haloes occupying more anisotropic matter environments.
          
          \item Tidal anisotropy affects OSB and ASB differently. Matching the tidal anisotropy $\alpha$ suppresses much of OSB over the mass range studied, but it has little effect on ASB. Thus tidal anisotropy can explain OSB in part, but not the relation between halo orientation and secondary halo properties.
          
          \item Halo definition is a significant systematic for spin OSB but a minor one for ASB. Consistent with \cite{2026Ma_UDGspin}, including unbound particles affects the low-mass spin-bias inversion; however, the orientation-dependent ASB trends are nearly unchanged across the halo definitions tested here.
     \end{itemize}
     
     These results distinguish four related but distinct manifestations and systematics: halo-environment alignment traces and partly absorbs the orientation dependence of spin and shape, outer matter anisotropy does not by itself generate ASB, tidal anisotropy controls part of the OSB signal, and halo definition primarily affects spin bias at low mass. Together, they fit within the same anisotropic large-scale-structure formation picture. Accounting for these distinctions will be important for using halo secondary properties in galaxy-halo models and for assessing orientation-dependent systematics in clustering, intrinsic-alignment, and redshift-space-distortion analyses.
     
     %% Also note that the akcnowlodgment environment does not support long amounts of text. If you have a lot of people and institutions to acknowledge, do not use this command. Instead, create a new \section{Acknowledgments}.
     \begin{acknowledgments}
     This work is supported by the National Natural Science Foundation of China (grant No. 12433003).
     \end{acknowledgments}
     
     \section*{Data Availability}
     
     The IllustrisTNG data products used in this work are publicly available at \url{https://www.tng-project.org/data/}. Derived catalogues and analysis scripts used to make the figures will be shared by the corresponding author upon reasonable request.
     
     %% To help institutions obtain information on the effectiveness of their 
     %% telescopes the AAS Journals has created a group of keywords for telescope 
     %% facilities.
     %
     %% Following the acknowledgments section, use the following syntax and the
     %% \facility{} or \facilities{} macros to list the keywords of facilities used 
     %% in the research for the paper. Each keyword is check against the master 
     %% list during copy editing. Individual instruments can be provided in 
     %% parentheses, after the keyword, but they are not verified.
     
     %\vspace{5mm}
     %\facilities{HST(STIS), Swift(XRT and UVOT), AAVSO, CTIO:1.3m,
     %CTIO:1.5m,CXO}
     
     %% Similar to \facility{}, there is the optional \software command to allow 
     %% authors a place to specify which programs were used during the creation of 
     %% the manuscript. Authors should list each code and include either a
     %% citation or url to the code inside ()s when available.
     
     %% Appendix material should be preceded with a single \appendix command.
     %% There should be a \section command for each appendix. Mark appendix
     %% subsections with the same markup you use in the main body of the paper.
     
     %% Each Appendix (indicated 16
     %with \section) will be lettered A, B, C, etc.
     %% The equation counter will reset when it encounters the \appendix
     %% command and will number appendix equations (A1), (A2), etc. The
     %% Figure and Table counter will not reset.
     
\appendix

     \renewcommand{\thefigure}{A\arabic{figure}}
     \renewcommand{\theHfigure}{A\arabic{figure}}
     \renewcommand{\theHequation}{A\arabic{equation}}
     \setcounter{figure}{0}

     \section{Halo-environment alignment and outer matter anisotropy}
     \label{app:matter_anisotropy}

     This appendix collects supporting tests for the environmental alignment and outer matter anisotropy measures defined in Section \ref{data:anisotropy}. To keep the appendix focused, we show the outer-matter geometry. The representative OSB control for $c/a(4R_{\rm vir})$ is now included in Figure \ref{fig:matteraniso_all}. We have also checked the corresponding controls with the outer-matter/tidal alignment $|\mathbf{A}_{4R_{\rm vir}}\cdot\mathbf{e}_3|$; they give the same qualitative conclusions and are not shown separately.
     
\begin{figure*}[htpb]
       \includegraphics[width=1.0\textwidth]{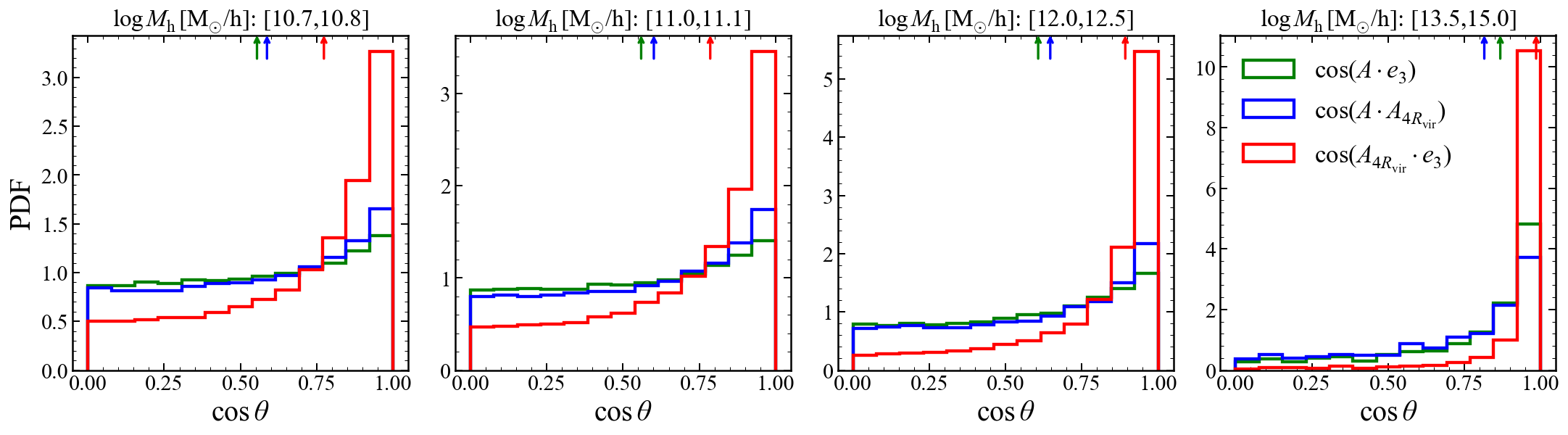}
       \caption{Probability distributions of alignment cosines in four halo-mass ranges. Green histograms show $|\mathbf{A}\cdot\mathbf{e}_3|$, where $\mathbf{A}$ is the halo major axis and $\mathbf{e}_3$ is the slowest-collapse direction of the tidal field. Blue histograms show $|\mathbf{A}\cdot\mathbf{A}_{4R_{\rm vir}}|$, where $\mathbf{A}_{4R_{\rm vir}}$ is the major axis of dark matter in the shell $2R_{\rm vir}<r<4R_{\rm vir}$. Red histograms show $|\mathbf{A}_{4R_{\rm vir}}\cdot\mathbf{e}_3|$. Arrows mark the median values of the corresponding distributions.}
       \label{fig:outer4R_alignment}
     \end{figure*}

     Figure \ref{fig:outer4R_alignment} shows how the halo major axis, outer matter major axis, and tidal-field direction are mutually aligned. The outer matter major axis $\mathbf{A}_{4R_{\rm vir}}$ is strongly aligned with $\mathbf{e}_3$, especially at high mass. The halo major axis $\mathbf{A}$ is also preferentially aligned with both $\mathbf{A}_{4R_{\rm vir}}$ and $\mathbf{e}_3$, but with broader distributions. This confirms that the outer matter shape traces the large-scale filamentary direction, while also showing why halo-axis alignment and matter anisotropy amplitude should be treated separately.
     
     \renewcommand{\thefigure}{B\arabic{figure}}
     \renewcommand{\theHfigure}{B\arabic{figure}}
     \renewcommand{\theHequation}{B\arabic{equation}}
     \setcounter{figure}{0}

     \section{\texorpdfstring{Halo-definition robustness}{Halo-definition robustness}}
     \label{app:halo_def}

\begin{figure*}[htpb]
       \includegraphics[width=1.0\textwidth]{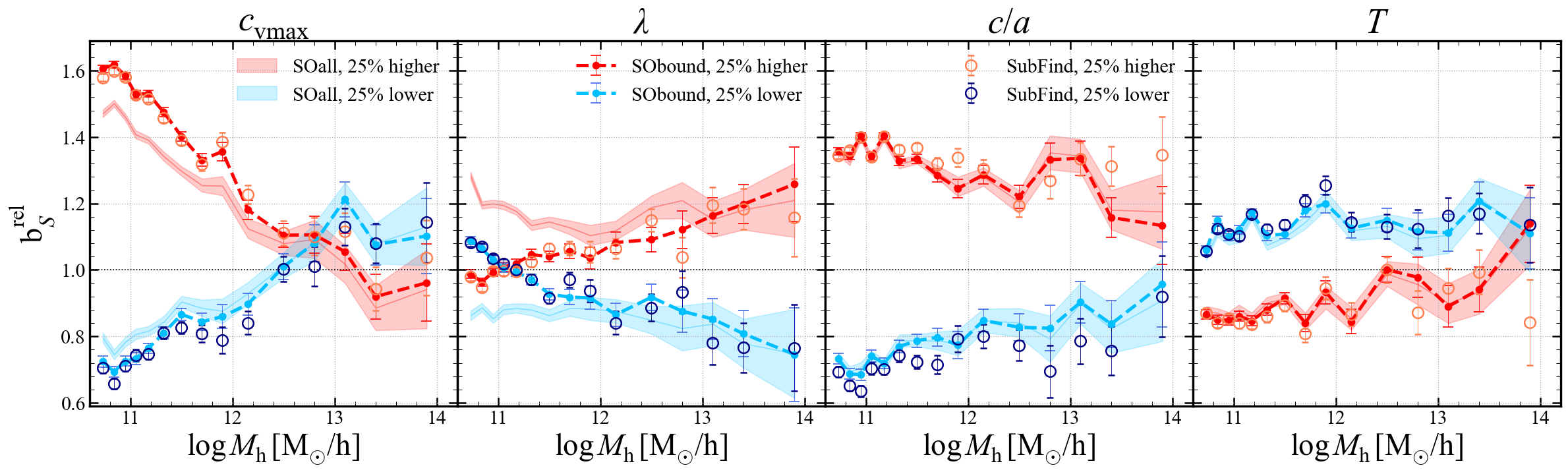}
       \caption{Halo-definition comparison of OSB. Columns, from left to right, show concentration $c_{vmax}$, spin $\lambda$, minor-to-major axis ratio $c/a$, and triaxiality $T$. The upper panels show the absolute bias of all haloes and of the upper and lower 25\% subsamples. The lower panels show the corresponding relative bias $b^{rel}_{S}$. Solid curves with shaded bands show SOall, filled circles with error bars show SObound, and open circles with error bars show SubFind. Red and blue denote the upper and lower 25\% subsamples. The horizontal dotted line in the lower panels marks $b^{rel}_{S}=1$.}
       \label{fig:multihalopro_diff_def}
     \end{figure*}

     Figure \ref{fig:multihalopro_diff_def} illustrates the halo-definition dependence of OSB for concentration, spin, shape, and triaxiality. The signal is strongest for spin, where the low-mass spin-bias inversion is present for SObound but absent for SOall. SubFind gives a similar low-mass inversion to SObound, indicating that the inversion is mainly associated with excluding unbound particles. Concentration shows a smaller definition dependence at low mass, while shape and triaxiality are nearly identical for all three definitions.
     
\begin{figure*}[!htpb]
       \includegraphics[width=1.0\textwidth]{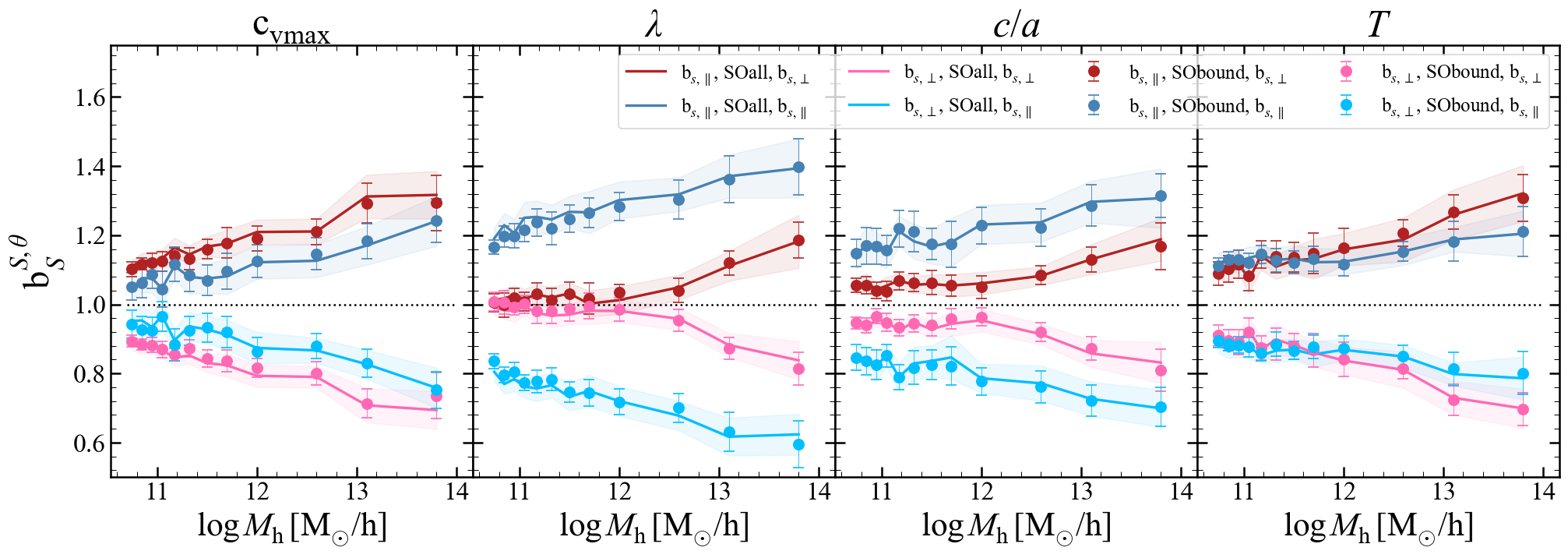}
       \caption{Anisotropic secondary bias for the SOall and SObound definitions. Columns, from left to right, show concentration $c_{vmax}$, spin $\lambda$, minor-to-major axis ratio $c/a$, and triaxiality $T$. The plotted quantity is $b^{S,\theta}_{S}$ as a function of halo mass. Solid curves with shaded bands show SOall, and filled circles with error bars show SObound. Dark red and dark blue denote the upper and lower 25\% subsamples measured parallel to the halo major axis, while pink and cyan denote the corresponding upper and lower 25\% subsamples measured perpendicular to the halo major axis. The horizontal dotted line marks $b^{S,\theta}_{S}=1$.}
       \label{fig:secb_theta_dissh}
     \end{figure*}

     Figure \ref{fig:secb_theta_dissh} compares the ASB measurements obtained with the SOall and SObound definitions. The comparison isolates the effect of unbound particles while keeping the same spherical-overdensity halo boundary. The two definitions give similar orientation-dependent trends for concentration, spin, shape, and triaxiality, supporting the conclusion that halo definition affects spin OSB more strongly than the main ASB signal.

     %Appendices can be broken into separate sections 
     %% For this sample we use BibTeX plus aasjournals.bst to generate the
     %% the bibliography. The sample631.bib file was populated16 from ADS. To
     %% get the citations to show in the compiled file do the following:
     %%
     %% pdflatex sample631.tex
     %% bibtext sample631
     %% pdflatex sample631.tex
     %% pdflatex sample631.tex
     
\bibliography{sample701}{}
\bibliographystyle{aasjournalv7}

%% This command is needed to show the entire author+affiliation list when
%% the collaboration and author truncation commands are used. It has to
%% go at the end of the manuscript.
%\allauthors

\end{document}